\documentclass{article}

\usepackage{arxiv}

\usepackage[utf8]{inputenc} 
\usepackage[T1]{fontenc}    
\usepackage{hyperref}       
\usepackage{url}            
\usepackage{booktabs}       
\usepackage{amsfonts}       
\usepackage{nicefrac}       
\usepackage{fontawesome}
\usepackage{microtype}      
\usepackage{lipsum}
\usepackage{graphicx}
\usepackage{amsmath}
\usepackage{algorithm}
\usepackage{algpseudocode}
\graphicspath{ {./Images/} }

\title{Fair Compromises in Participatory Budgeting: a Multi-Agent Deep Reinforcement Learning Approach}

\author{
 Hugh Adams \\
  School of Computer Science\\
  University of Leeds\\
  Leeds, UK LS2 9JT \\
  \texttt{schad@leeds.ac.uk} \\
   \And
 Srijoni Majumdar \\
  School of Computer Science\\
  University of Leeds\\
  Leeds, UK LS2 9JT \\
  \texttt{s.majumdar@leeds.ac.uk} \\
  \And
 Evangelos Pournaras \\
  School of Computer Science\\
  University of Leeds\\
  Leeds, UK LS2 9JT \\
  \texttt{e.pournaras@leeds.ac.uk} \\
}

\begin{document}
\maketitle
\begin{abstract}
Participatory budgeting is a method of collectively understanding and addressing spending priorities where citizens vote on how a budget is spent, it is regularly run to improve the fairness of the distribution of public funds. Participatory budgeting requires voters to make decisions on projects which can lead to ``choice overload". A multi-agent reinforcement learning approach to decision support can make decision making easier for voters by identifying voting strategies that increase the winning proportion of their vote. This novel approach can also support policymakers by highlighting aspects of election design that enable fair compromise on projects. This paper presents a novel, ethically aligned approach to decision support using multi-agent deep reinforcement learning modelling. This paper introduces a novel use of a branching neural network architecture to overcome scalability challenges of multi-agent reinforcement learning in a decentralized way. Fair compromises are found through optimising voter actions towards greater representation of voter preferences in the winning set. Experimental evaluation with real-world participatory budgeting data reveals a pattern in fair compromise: that it is achievable through projects with smaller cost. 
\end{abstract}


\section{Introduction}
Participatory budgeting is a collective decision making process where citizens vote on how a budget is spent; distribution of resources is informed by public opinion rather than by government officials. 
It is seen as a way to improve financial efficiency \cite{pournaras2025upgradingdemocraciesfairervoting}, to better understand the demands of the public and to address them more fairly. 

Voting in participatory budgeting elections is often demanding on participants. They must learn about and form preferences on an array of projects, as well as decide how to express these preferences in complex ballot formats. In the Stadtidee election in Aarau, Switzerland there were 33 projects ranging from a public herb garden to a children's disco \cite{pournaras2025upgradingdemocraciesfairervoting}. Too many alternative projects can overwhelm voters, causing ``choice overload" \cite{Iyengar2000}. Voters are also faced with the problem that the efficacy of their vote is dependent on how every other participant votes; participants often cast their vote strategically based on their preferences and their expectations of possible outcomes \cite{Giorgetta2021}. Artificial intelligence based decision support could assist populations to find fair compromises, adopting a hybrid intelligence \cite{Dellermann2019} approach to the problem of efficiency and fairness in budgeting. This support could work for two groups; it could help voters build more accurate understanding of alternative outcomes and point to routes for favourable compromises; and for policymakers, it could aid election design by highlighting possible collective choices that improve fairness.


Artificial intelligence decision support, especially for voting, needs careful consideration of ethics. Models are by definition reductive. What is useful about a model is that reductions in complexity can still capture what is important about the model while making it understandable. Since a model can not capture everything about a system, it at best captures the system accurately from one perspective. This perspective is defined by the modeller in what is featured in the model and what assumptions are made. So a first point to ethically align decision support is to be transparent about model construction, being clear what assumptions were made and why. In this paper this especially applies to assumptions around voter decision making, an assumption that is left out of related literature and often taken for granted.

A second point is that when models are able to pick up on patterns, these are historically contingent patterns of behaviour, not transhistorical truths \cite{10.1162/artl_a_00336}. So, when models of human behaviour make predictions of future behaviour, this ties people to their past. Even when models accurately pick up historical patterns in the data, care needs to be taken in how these are incorporated into supporting decision making. A simple recommendation approach, where an AI system makes recommendations of actions to voters based identifying voters historical preferences can discount the capacity for people to change and grow. What models are good at is playing with what-ifs. By conducting experiments on models, relations between model elements can be drawn out:``often the main benefit of designing and using a model is that it provides an understanding of the policy domain, rather than the numbers it generates'' \cite{gilbert2018}. So in this paper, the model experiments can serve as decision support, not by providing individual recommendations to each voter, but by highlighting general insights (opting for lower costing projects) that could provide a route to fair compromises in the election contexts covered.

This work proposes a multi-agent reinforcement learning approach to decision support, presenting possible collective choices from optimised actions. In Reinforcement learning, goals and decision making criteria are specified in the reward function \cite{sutton2018reinforcement} and behaviour is learnt through optimising with respect to the reward function. This makes reinforcement learning a useful method for decision support, alternative actions and collective choices are discovered through optimisation.

A compromise is an altered action not seen as ideal, made in recognition that the ideal may not be achievable. In most current literature modelling elections with multi-agent reinforcement learning, voters have a complete internal ordering of preferences for projects from which they select an action. This leaves preference formation outside of the model. Voters form opinions on projects as they encounter them, and they must form these preferences based on some higher level world-view or preference criteria. To investigate a route to fair compromise, this work includes a framework for voter preference formation based on ``issue voting'' model of voter behaviour \cite{DENVER199019}: voters favour projects based on their cost and contributions to wider issues. The framework used in this work is a robust starting point, intended to open up investigation of higher-order preferences and how they are expressed given a voting context and their relationship to fair compromise. By starting with internal preference ordering, it is only possible to ask ``\textit{how much}'' voters are willing to compromise, by including higher-order preference formation it is possible to ask ``\textit{what}'' voters are willing to compromise on, such as project cost or projects that support a particular issue.

The experiments presented in this work consist of modelling two participatory budgeting elections, the "Stadtidee" election in Aarau, Switzerland 2023 \cite{pournaras2025upgradingdemocraciesfairervoting}, and the Budget "Participatif" in Toulouse, France 2019. The multi-agent reinforcement learning models are then contrasted to the actual elections in terms of voting behaviour and fairness of collective choice. These models point to a fairer, more satisfying collective choice through compromising on high cost  of projects. A significant methodological challenge of these experiments was designing a reinforcement learning algorithm for voting agents that could scale. The combinatorial nature of the ballots used lead to exploding action spaces. This challenge is overcome by utilising an action branching architecture with deep Q-learning

The contributions of this paper are as follows: (i) a new multi-agent reinforcement learning approach to modelling participatory budgeting. (ii) a novel solution for decentralised voting agent decision making at scale: action branching deep Q-learning. (iii) a route to fair compromise identified through multi-agent reinforcement learning modelling: a shift to lower cost projects can promote fairness and higher collective voter satisfaction.

This paper is outlined as follows; Section \ref{Related work} reviews related work; Section \ref{Methods} introduces the participatory budgeting model and agent decision making model; Section \ref{Experimental methodology} outlines the experiment parameters and set up; Section \ref{Results} illustrates the findings of the experiments; Section \ref{Discussion} concludes the paper and outlines future research directions.

\section{Related Work}\label{Related work}
This section provides an overview of the related literature, with a focus on multi-agent reinforcement learning modelling of collective choice procedures, a summary of which can be found in Table~\ref{tab:comparison}. Related work is contrasted in terms of; whether it is modelling participatory budgeting or single choice voting; whether voter preference formation is included in model scope; whether it utilises real world data; whether voters are assumed to be self interested; how well it scales.

Both Liekah \& Grandi \cite{liekah2019multiagent}, and Airiau et al. \cite{airiau} utilise multi-agent reinforcement learning for modelling voters in a voting game. Their agents start with a preference ordering of states and choose which to select. Their work demonstrates learning agents' capacity to reach ``good'' collective choice; however, they do not investigate what types of compromises were taken to arrive at these results. One framework for voter preference formation is projects' contributions to wider issues or impact areas, such as ``culture", or ``welfare" \cite{maharjan2024fairvotingoutcomesimpact}. Majumdar \& Pournaras' \cite{majumdar2023consensusbased} MARL-PB uses this impact area formulation to demonstrate consensus as a viable aggregation method for participatory budgeting. However, they do not analyse which aspects of projects make up these compromises.

Mertzanie et al. use multi-agent reinforcement learning to illustrate a novel consensus process: \(\theta\)-learning \cite{thorybosLearning}. In their paper, they model the ancient Athenian process "Thorybos". Modelling an abstract situation can be useful when illustrating a new or complicated idea, such as "Thorybos". However, for supporting decision making, predictions need to be falsifiable \cite{edmonds2019}. This means that for this purpose models should refer to a specific context and incorporate data from that context in the model.

Multi-agent reinforcement learning suffers from the problem of non-stationarity, the environment changes with regard to the actions of all agents \cite{marl-book}. Reward functions are often extended to improve convergence, such as the addition of a collective reward  \cite{thorybosLearning}, or the reward for communication \cite{majumdar2023consensusbased}. These extensions come with assumptions about voter motivations. A collective reward means that an agent's goal also includes the general well-being of the population, the inter-agent communication reward assumes that a ballot seems more appealing the more it is recommended. Both of these assumptions are justifiable empirically. However, there is value in using an unextended, purely self-interested, reward function for modelling fair compromise. The proposed compromise cannot be critiqued in terms of not being in the interest of voters.

One challenge of multi-agent reinforcement learning is scalability: the more voting agents, the more the computational cost. Of the previously mentioned comparable papers, the maximum number of agents in the models are 9 \cite{airiau}, 15  \cite{liekah2019multiagent}, and 100 \cite{majumdar2023consensusbased}. The maximum number of agents in experiments in this paper is 1703. One cause of the scalability problem in modelling participatory budgeting is the ballot format. Compared to a simple choice of one candidate in general elections, participatory budgeting ballots often contain multiple candidates. This paper uses a branching architecture deep Q-learning algorithm \cite{BDQ}, where the neural network deals with the action space with multiple output heads. 
\begin{table*}[!htb] \caption{Research Gap}\label{tab:comparison} 
\centering\resizebox{0.75\paperwidth}{!}{%
\begin{tabular}{p{1.7in}p{0.9in}p{0.9in}p{1.1in}p{1.2in}p{0.7in}}\toprule
\textbf{ Method feature}& liekah et al. \cite{liekah2019multiagent}  &Airiau et al. \cite{airiau} & Mertzanie et al. \cite{thorybosLearning}  & Majumdar et al. \cite{majumdar2023consensusbased} & \textbf{This Paper}\\\cmidrule(lr{.5em}){2-6}
 \addlinespace
 \textbf{Participatory budgeting} & \faCheck & \faTimes  & \faTimes &  \faCheck   & \faCheck\\
 \textbf{Preference Modelling}& \faTimes & \faTimes & \faTimes & \faCheck  & \faCheck \\
  \textbf{Real-world data}& \faTimes & \faTimes &  \faTimes & \faCheck & \faCheck  \\
 \textbf{Self-interested rewards}&  \faCheck & \faCheck & \faTimes & \faTimes &\faCheck\\
 \textbf{Scalability}& \faTimes &\faTimes &  \faTimes  & \faTimes  & \faCheck\\
\bottomrule
\end{tabular}%
}
\end{table*}

\section{Methods}
\label{Methods}
\subsection{Participatory Budgeting Model}

\begin{figure*}[h!]
    \centering
    \includegraphics[width=0.7\textwidth]{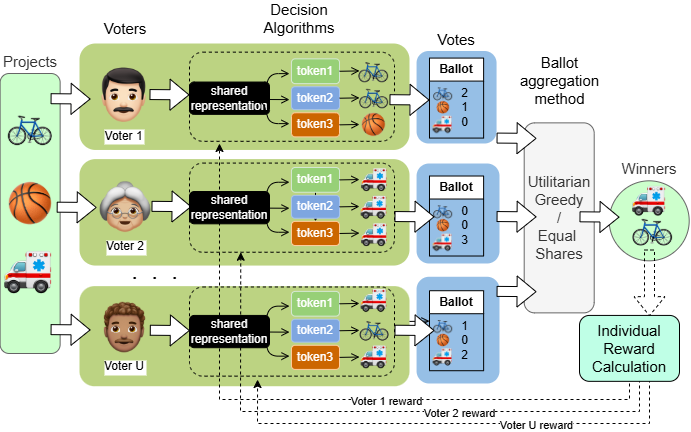}
    \caption{\textbf{A conceptual visualisation of the objects and flow of one episode of the participatory budgeting model.} Projects are presented to a population of voting agents. These agents then utilise their decision module to produce cumulative ballots by separating each token assignment to a parallel output. These votes are then aggregated to produce winners. An individualised reward is returned to voters. }
    \label{fig:sivo}
\end{figure*}
Figure~\ref{fig:sivo} shows a conceptual visualisation of the participatory budgeting model used in this paper. The components that make up the model are; voting agents; projects; impact areas; budget; ballot format; votes; ballot aggregation method and the set of winning projects. An impact area determines the scope and relevance of a project in society, eg `environmental protection'. A project is a fully costed and complete plan, such as a new hospital. A project comes with a cost and the impact areas it contributes to. The budget of an election is the maximum amount of money that can be spent on the projects in the winning set. The ballot aggregation method is a function that takes the full profile of votes and the budget, and it outputs a winning set of projects. The winning set of projects is a subset of the full set of projects that adheres to the budget.

\paragraph{Ballot Format} 
Cumulative ballots give each voter a number of tokens they can assign to projects. Suppose there are 10 projects and 5 tokens to each cumulative ballot, a voter could assign 5 tokens to one project, or one token to 5 projects, or 3 tokens to one project and 2 tokens to another. Cumulative voting is known for giving a stronger voice for minorities \cite{bhagat1984cumulative}. In the actual election in Aarau, there is an additional condition on the ballot that tokens must go to at least 3 projects if 3 or more tokens are assigned. While actual votes were cast in this manner, this condition was relaxed in the reinforcement learning model to simplify the action space. 

\paragraph{Ballot Aggregation}
The utilitarian greedy recursively selects the most popular project that is within the available budget. Method of equal shares is a voting rule designed to produce fairer outcomes for participatory budgeting \cite{NEURIPS2021_69f8ea31}. The idea is that each voter is assigned an equal part of the budget to spend. Since equal shares often leaves a significant portion of the budget unspent, the rest of the budget is spent with utilitarian greedy or another termination strategy. Utilitarian greedy is the aggregation rule originally used in Toulouse, equal shares is the rule originally used in Aarau.

\subsection{Voting Agent Model}

Agents make actions (submit ballots) based on their policies (understanding of how valuable each action is to them) and their observation (projects and their associated costs and impact area contributions).

\paragraph{Algorithm}
The agent voting algorithm in this work is a branching variant of deep Q-network (DQN) \cite{Mnih2015HumanlevelCT}. In deep Q-learning, the policy is composed of a neural network that approximates the optimal Q function (a function that assigns values to an action in a given state). The policy function is greedy: The action with the highest Q value is selected. This is a parameterised policy, the Q values are dependant on the values of the weights and biases of the neural network which are updated with respect to the reward function in training. It is through training that these weights change, so that behavioural policy is optimised towards the decision making criteria encoded in the reward function. Each voting agent has its own distinct Q network, each voting agent learns different action policies over training. 

\begin{figure*}[h!]
    \centering
    \includegraphics[width=0.8\textwidth]{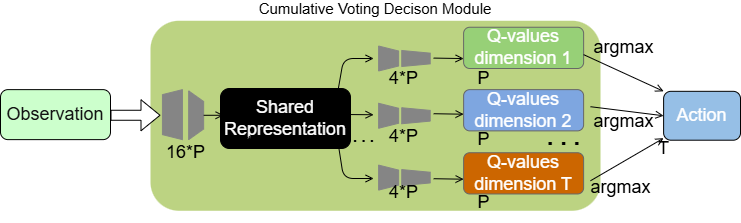}
    \caption{A visualisation of the branching architecture of a voting agent decision module. P= No. of Projects, T =  No. of Tokens}
    \label{fig:BDQ}
\end{figure*}

Branching neural network architectures have been shown to deal well with large action spaces \cite{BDQ}. A visualisation of the algorithm architecture is presented in Figure~\ref{fig:BDQ}. Cumulative ballots, the ballots of the elections featured in this paper, require voters to assign T tokens across P projects. So, the action space is all possible assignments of the T tokens across all projects. Branching splits the action space, from one space of T token assignments, to T spaces of 1 token assignment. The algorithms architecture branches out from a shared representation section composed of fully connected layers into T parallel action dimensions, each composed of fully connected layers. The grey trapeziums in Figure \ref{fig:BDQ} represent the fully connected neural network layers and the size of each layer is indicated in the number beneath. Each action dimension outputs P Q-values for a token assignment. The action with the largest Q-value is selected with argmax for each dimension and then are combined to form the action length T. The number of times a project appears in the vector is its score in the cumulative ballot. For example, in the case of Aarau \cite{pournaras2025upgradingdemocraciesfairervoting}, each voter has 10 tokens to assign across 33 projects. Without a branching architecture, the action space is \( \binom{33+10-1}{10} = 1,471,442,973\). With the branching architecture the action space is \(33 * 10 = 330\).

Experience replay is used to make experiments quicker and less computationally expensive \cite{af14a67b70be435f9657a304976993fe}. To perform experience replay, each agent's experiences, their state, action, and reward, $(s_t, a_t, r_t)$ are stored in an agent specific buffer. During learning updates, a sample of experience transitions, called a mini-batch, is obtained from the agent's buffer for training. The current network is updated with the following loss function
\(\
L = \mathbb{E}_{(s,a,r)  \sim D} [\frac{1}{N} \sum_d(r-Q_d(s,a_d))^2],
\)
where \(\mathbb{E}\) is the average operator, $(s,a,r)$ is a transition comprised of a state, an action and a reward, $D$ is the agent's buffer, \(N\) is the number of action dimensions, $Q_d$ is the Q-network for that agent, \(Q_d(s,a_d)\) is the Q-value of action \(a_d\) in state \(s\). Since both experiments consider a single election with fixed projects, the reward function has no future discount component. 

\paragraph{Reward Function}

\begin{figure*}[h!]
    \centering
    \includegraphics[width=0.8\textwidth]{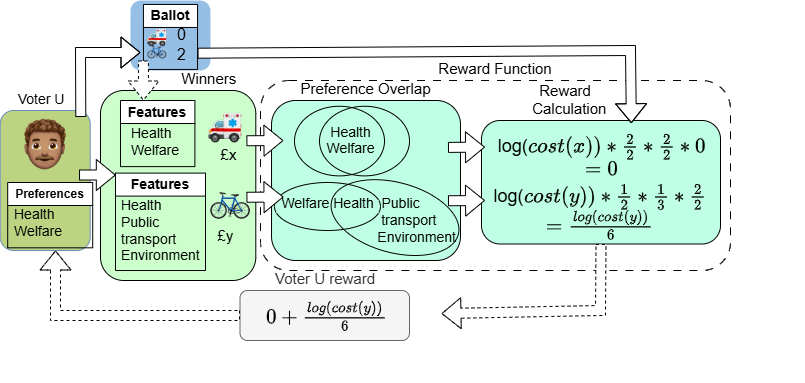}
    \caption{A visualisation of reward calculation: voters get a reward when they vote for a project that they favour that wins.}
    \label{fig:reward_fnnnn}
\end{figure*}

The reward function, illustrated in Figure \ref{fig:reward_fnnnn}, is an encoding of the decision making criteria of the voters. Voters favour impact areas, wider societal issues such as 'environmental protection' or 'public transport', and they value projects based on the project's contributions to these impact areas. The extent to which they favour projects is then related to the size of the project, this is measured as a function of the cost.  Formally, the reward function is defined as: 
\begin{equation}
\footnotesize
R(v,W) = \sum_{w \in W}\left( log(C(w)) \times \frac{p(v) \cap i(w)}{p(v)} \times \frac{p(v) \cap  i(w)}{i(w)} \times \frac{a(w)}{T}\right)
\end{equation}
where \(W\) is a winning set of projects, \(v\) is a voter, \(C(w)\) is the cost of project w, \(p(v)\) is the set of impact areas that voter v favours, \(a(w)\) is the number of tokens assigned to the project \(w\), T is the total number of tokens a voter can assign in a ballot, and \(i(w)\) is the set of impact areas that project w contributes to. Note that if there is no overlap between the preferences of a voter, and the contributions of a project from the winning set then the voter receives no reward from that project. Also note, that since the rewards are dependant on projects winning, the agents learn to value actions differently when they are trained with different ballot aggregation methods. Rewards are also dependant on a voter favouring that project, so if an agent votes for a project and that project wins, if the voter has no preference overlap, the voter receives no reward from that project winning.

\paragraph{Training}
Algorithm~\ref{alg:pseudocode} outlines the training process for the voters' decision modules. Neural network weights are initialised with Xavier initialisation, a technique for improving training stability and efficiency \cite{pmlr-v9-glorot10a}. Epsilon-greedy is chosen as the exploration strategy because of its robust performance with Q-learning. A mini-batch of transitions stored in an agent specific buffer is sampled with a selection that prioritises recent experiences \cite{DBLP:journals/corr/SchaulQAS15}.  Half of the mini-batch is composed of recent experiences (recent means transitions within the last 32 samples, 32 because it is the same as the batch size), the other half of the mini-batch is sampled from the rest of the transitions in the buffer.

\begin{algorithm*}
\scriptsize
\caption{Voting agent training.}\label{alg:pseudocode}
\begin{algorithmic}[1]
\State Initialise empty buffer and policy with random weights for each agent
\State Initialise election instance
\For{$episode = 1, M$}
    \State Each agent selects action with $\epsilon$-greedy
    \State Execute action profile to get winning set and rewards for each agent
    \State Store transitions in agents' buffers

    \If{training episode}
        \State For each agent, sample mini-batch of transitions from buffer
        \State For each agent, update Q
    \EndIf
\EndFor
\end{algorithmic}
\end{algorithm*}

\paragraph{Computational complexity}
For a DQN algorithm, the computational complexity is $\mathcal{O}(N((A + \chi) W + hW^2)) $ \cite{OMONIWA2023100640}, where N is the number of episodes, W is the number of nodes for a layer, A is the action space, $\chi$ is the state space and h is the number of hidden layers. For a branching algorithm, the computational complexity is $\mathcal{O}(N((BA + \chi) W + (r + Bb)W^2))$ Where B is the number of branches, r is the number of hidden layers in the shared representation layer, and b is the number of hidden layers of each branch.

The action space of the branched architecture is 33 for each branch, approximately  $3*10^2$, whereas the action space for the unbranched architecture is approximately $1.5*10^{9}$. Out of the experiments in this work, the maximum number of branches is 10. If the number of weights per layer is kept the same, the W term is an order of $10^6$ higher for the unbranched algorithm, and the $W^2$ term is order of $10^1$ lower for the unbranched algorithm. Overall, the branched algorithm achieves better complexity than the unbranched algorithm.

For a multi-agent problem with independent learning algorithms, the complexity scales proportionally with the number of agents, P. So computational complexity  for the experiments in this paper is $\mathcal{O}(NP((BA + \chi) W + (r + Bb)W^2))$

\section{Experimental Methodology}\label{Experimental methodology}
The aim of these experiments is to demonstrate a route to fairer collective choice through `learning' to make viable compromises. This is done by modelling past elections with Branching Q-learning algorithms for voting agents, utilising a decision making criteria  based on impact area contributions. The votes and collective choice of these models are then compared to the votes and collective choice from the actual election data. The collective choice is compared in terms of fairness, the votes are compared in terms of satisfaction and type of project voted for. These are also compared to the votes and collective choice of untrained agents, essentially random selections, to establish what it is that the agents learn.

\subsection{Election Datasets}

The elections used for the model in this work are the participatory budgeting elections of City Idea, Aarau, Switzerland from 2023 \cite{pournaras2025upgradingdemocraciesfairervoting}, the first participatory budgeting election to use cumulative voting and the method of equal shares \cite{NEURIPS2021_69f8ea31}, and the Budget participatif 2019 election in Toulouse, France \cite{DBLP:journals/corr/abs-2012-06539}. 

The Aarau election data was collected by the Trustworthy Distributed Intelligence research group at the University of Leeds \cite{pournaras2025upgradingdemocraciesfairervoting}. The data comes in a .pb file that is compatible with PABULIB \cite{DBLP:journals/corr/abs-2012-06539}. There are 1703 voters, 33 projects, and 9 impact areas. The Toulouse election data can be found on PABULIB \cite{DBLP:journals/corr/abs-2012-06539}. There are 1494 voters, 30 projects, and the same 9 impact areas. The impact area labels for the proposed projects were absent in the data. As such, the impact areas labels for proposed projects in these instances were assigned independently by multiple members of the Trustworthy Distributed Intelligence research lab to cross-validate the classification. Voter preferences for categories are abstracted from the votes that voters cast in the data: voters are assumed to favour all impact areas of the projects they originally voted for. This is a crude assumption, as voters might have voted for a project in spite of an impact area contribution. However, given the limited data, this assumption can be built on in future work.

\begin{figure*}[h!]
    \includegraphics[width=1.0\textwidth]{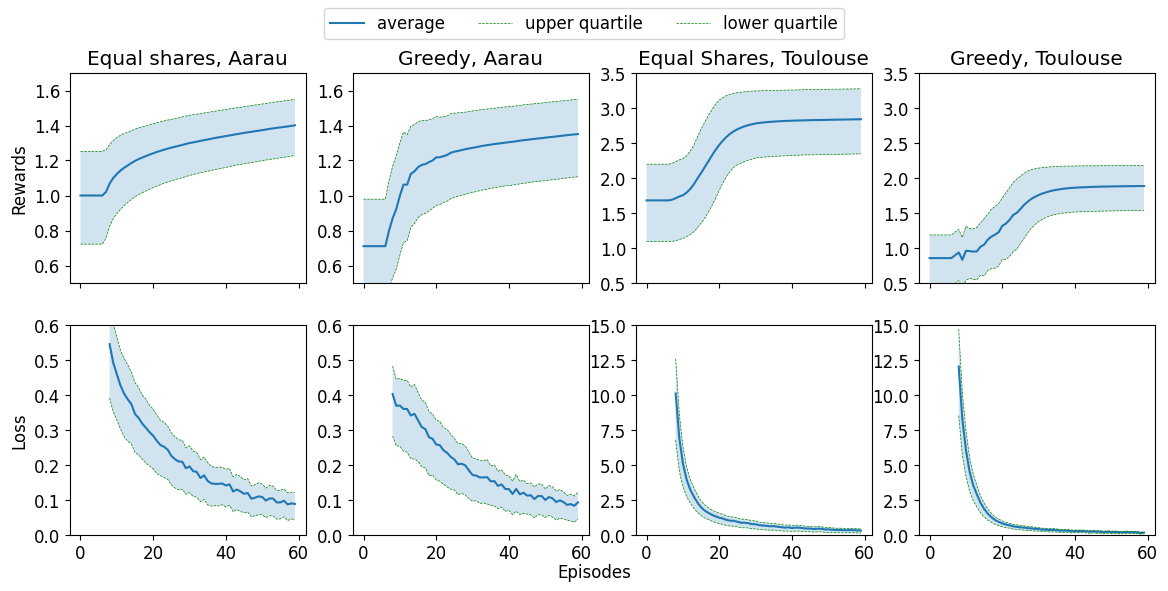}
    \caption{\textbf{Training increases accumulated rewards and reduces accumulated loss.} The central thick blue line represents the average loss or reward taken over every agent in every experiment for that validation epoch. The dotted green lines at the edge of the envelope show the inter-quartile range.}
    \label{fig:training}
\end{figure*}
\subsection{Experimental Parameters}
Experiments were conducted with the parameters outlined in Tables~\ref{tab:experimentparams} and~\ref{tab:modelparams}. Across every experiment, agents improve their reward with training and reduce their loss as shown in 
Figure \ref{fig:training}. Data from experiments illustrating results are taken from validation episodes. Every five training episodes, a validation episode is run, where actions are selected with optimal action, and the loss is not used to train the neural networks. The validation interval is the number of training episodes between each validation episode. Learning parameters were chosen based on standard values or methods in the literature.

\begin{table}[!htb]
\centering
\begin{minipage}{0.49\textwidth}
\centering
\caption{Learning experiment hyperparameters}
\resizebox{\textwidth}{!}{%
\begin{tabular}{lrr}
\toprule
\textbf{Experiment parameters} & \multicolumn{2}{c}{Value} \\
\midrule
City & Aarau & Toulouse \\
Number of voters & 1703 & 1494 \\
Number of projects & 33 & 30 \\
Number of impact areas & 9 & 9 \\
Number of Ttkens & 10 & 7 \\
Repetitions & 10 & 10 \\
Budget & 50000 CHF & 1000000 EUR \\
\bottomrule
\end{tabular}
}
\label{tab:experimentparams}
\end{minipage}
\hfill
\begin{minipage}{0.49\textwidth}
\centering
\caption{Model learning and PB parameters}
\resizebox{\textwidth}{!}{%
\begin{tabular}{lr}
\toprule
\textbf{Learning parameters} & \\
\midrule
Validation interval & 5 \\
Learning rate & 0.001 \\
Batch size & 32 \\
Target net update frequency & 100 \\
Discount factor, $\gamma$ & 0 \\
Exploration & $\epsilon$-greedy \\
Training episodes & 400 \\
\midrule
\textbf{Participatory budgeting parameters} & \\
\midrule
Vote aggregation method & Equal shares/Greedy \\
Ballot format & Cumulative \\
\bottomrule
\end{tabular}
}
\label{tab:modelparams}
\end{minipage}
\end{table}

\subsection{Welfare Measures}
\textit{Satisfaction (projects)} \(rep_p=\frac{|a \cap W|}{|W|}\) where \(a\) is the action of the voting agent, the projects that it voted for, and W is the winning set of the election. 
\textit{Satisfaction (cost)} \(rep_c=\frac{c(a \cap W)}{c(W)}\) where \(a\) is the action of the voting agent, the projects that it voted for, W is the winning set of the election, and c is the cost function.
\textit{Share} \(\sigma=\sum_{p \in a \cap W} \frac{c(p)}{v(p)}\) where \(a\) is the action of the voting agent, the projects that it voted for, W is the winning set of the election, and c is the cost function, v is the number of votes that p received.

\subsection{Fairness Criteria}\label{fairness}
\textit{Egalitarian social welfare}
A measure of how well off the worst off in a society is. It is equal to the welfare of the voter with the lowest welfare \cite{10.1007/978-3-540-25946-6_6}. A non-zero egalitarian welfare indicates that every voters has some project they voted for in the winning set, indicating that every voter is included in the benefits of the election.
\textit{Utilitarian social welfare} 
A measure of how well off society is as a whole \cite{10.1007/978-3-540-25946-6_6}. It is equal to the sum total welfare of every citizen in the society. In this work it is divided by the number of voters in the election, so that it is on a similar scale with other fairness criteria.
\textit{Gini} The Gini coefficient is a measure of inequality \cite{ijcai2021p42}. 0 represents maximum equality and 1 maximum inequality.

\section{Results}\label{Results}
\textbf{\textit{Three key results are illustrated in this paper}} - (i) Voting agents in the proposed multi-agent reinforcement learning model achieve a higher vote satisfaction than the one of voters in the actual election - (ii) The collective choice produced with multi-agent reinforcement learning model is fairer than the collective choice in the actual elections - (iii) Improved fairness and vote satisfaction is achieved via compromise in voting choices which prioritises voting for projects with lower cost. 

\begin{figure*}[h!]
    \includegraphics[width=1.0\textwidth]{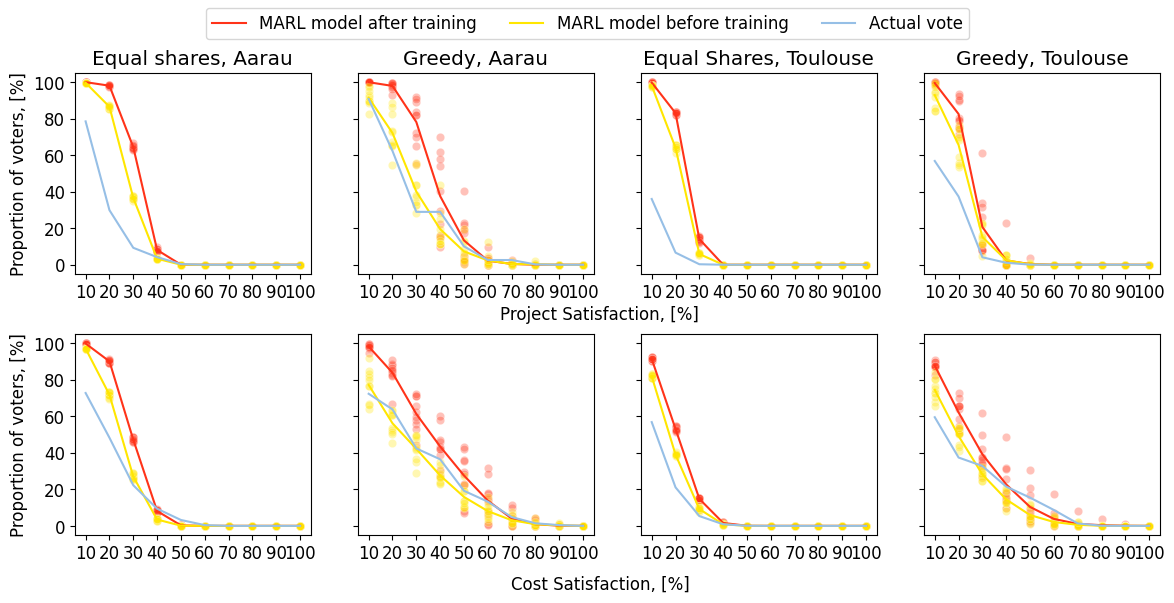}
    \caption{\textbf{The proportion of project and cost satisfaction of reinforcement learning voters after training is higher than the satisfaction before training and of the satisfaction of the voters in the actual election. } Satisfaction is measured in the proportion of voters out of the whole population. For example, in Equal shares, Aarau, 100\% of the trained MARL population achieves at least 10\% project satisfaction, and 0\% of the population achieves 60\% project satisfaction}
    \label{fig:satisfaction}
\end{figure*}

\paragraph{Voting agents in multi-agent reinforcement learning achieve a higher vote satisfaction than from voters in the actual election.}

Figure \ref{fig:satisfaction} shows that across Aarau and Toulouse, in both aggregation methods, the trained reinforcement learning agents have a higher project and cost satisfaction than the actual voters. In addition to this, the effect of training on the reinforcement learning agents is to improve their project and cost satisfaction.

\paragraph{The collective choice produced with multi-agent reinforcement learning model is fairer than the collective choice in the actual elections.}

\begin{table}[!htb] \caption{Collective choice fairness}
\label{tab:collective_choice_fairness}
\centering\resizebox{0.75\paperwidth}{!}{%
\begin{tabular}{l|cccccccccccc} 
&\multicolumn{4}{c}{Gini} & \multicolumn{4}{c}{Egalitarian} & \multicolumn{4}{c}{Utilitarian} \\
&\multicolumn{2}{c}{Actual} & \multicolumn{2}{c}{MARL} & \multicolumn{2}{c}{Actual} &\multicolumn{2}{c}{MARL}& \multicolumn{2}{c}{Actual} &\multicolumn{2}{c}{MARL} \\
& Equal shares &Greedy & Equal shares &Greedy & Equal shares &Greedy & Equal shares&Greedy & Equal shares &Greedy & Equal shares&Greedy \\\midrule
\textbf{Aarau} \\\hline
Satisfaction (project)&0.33&0.33&0.11&0.17&0.00&0.00&5.26&0.00&16.9&27.8&31.1&36.7\\ \hline
   Satisfaction (cost)&0.37&0.43&0.14&0.30&0.00&0.00&6.60&0.00&20.3&29.4&29.5&37.9\\ \hline
                 Share&0.37&0.42&0.15&0.30&0.00&0.00&4.22&0.00&28.8&29.1&31.2&24.5\\ \midrule
\textbf{Toulouse} \\\hline
Satisfaction (project)&0.33&0.34&0.09&0.12&0.00&0.00&8.70&5.26&9.85&14.6&23.6&25.0\\ \hline
   Satisfaction (cost)&0.43&0.54&0.21&0.32&0.00&0.00&1.20&0.60&11.85&21.6&21.2&27.8\\ \hline
                 Share&0.47&0.48&0.26&0.31&0.00&0.00&29.5&12.0&667&669&502&567\\
\bottomrule
\end{tabular}%
}
\end{table}

Table \ref{tab:collective_choice_fairness} compares fairness of collective choice between trained reinforcement learning agents in the election contexts of Aarau and Toulouse, for both equal shares and greedy. Across both elections and ballot aggregation methods, Gini is smaller for all three welfare metrics for the reinforcement learning experiment than for the actual vote. This means that there is a greater equality of satisfaction and share in the reinforcement learning experiments than in the actual elections in both Toulouse and Aarau. The middle column of Table \ref{tab:collective_choice_fairness} compares the egalitarian welfare. This is 0 for all actual elections, meaning that in Toulouse and Aarau, whether using greedy or equal shares, at least one voter had none of the projects they voted for win. The reinforcement learning experiment in Aarau using greedy does not improve on this, but in Toulouse the values are non-zero, both equal shares reinforcement learning experiments had non-zero values. For these experiments, every voter had some part of their vote win. The right-hand column of Table \ref{tab:collective_choice_fairness} shows the utilitarian social welfare. The utilitarian social welfare is larger for equal shares and greedy in Toulouse and Aarau, in both cost and project satisfaction. This means that in the reinforcement learning experiments the total satisfaction increases. However, the share decreases in the reinforcement learning experiments for both aggregation methods in Toulouse, and for greedy in Aarau. So, the amount of the budget spent on each voter decreases. 

\paragraph{Voting agents in multi-agent reinforcement learning attributed a higher proportion of tokens to lower costing projects than voters in the actual election.}
\begin{figure*}[h!]
    \includegraphics[width=1.0\textwidth]{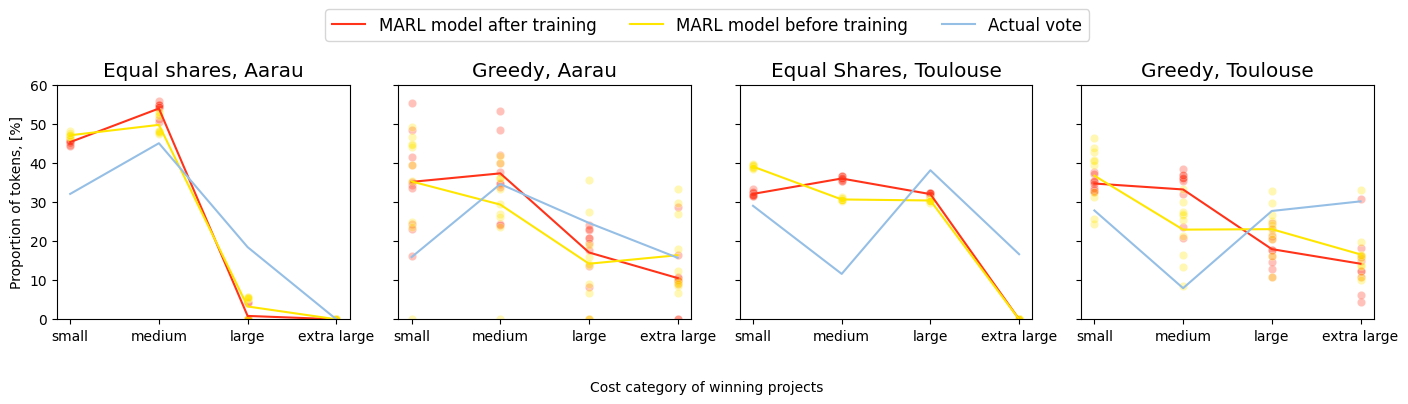}
     \caption{\textbf{Vote share distribution over project cost for Aarau and Toulouse.} The cost categories, small, medium, large and extra large, are segmented by quartile, so the cheapest 25 percent of projects are included in the small category. They are measured in the proportion of tokens assigned out of all possible tokens.}
    \label{fig:cost}
\end{figure*}

Figure \ref{fig:cost} compares the distribution of vote share over projects based on the size of their cost for the voting agent behaviour before and after training and for the voter behaviour in the actual election. Across both election contexts, reinforcement learning agents attribute a higher or equal token share to small and medium costed projects and a lower token share to large and extra large costed projects compared to the voters in the actual election.

\section{Discussion, Conclusion and Future Work}
\label{Discussion}

This work presents an approach to decision support through multi-agent reinforcement learning. A model is applied to two participatory budgeting elections to highlight routes to compromise. The results of these experiments indicate that a fairer more satisfying collective choice is possible through shifting votes to cheaper projects.

This work finds three main categories of beneficiaries; \emph{for research and design of decision support for elections}, the experiments are able to point towards behaviour changes favourable to voters because the reinforcement learning agent models include the process of project preference formation. This extension of the model scope allows research to be conducted on how voter motivations and decision making criteria relate to compromises and qualities of collective choice; \emph{for policymakers} modelling can inform the design of participatory budgeting elections to improve the fairness of collective choice, and so improve legitimacy of participatory budgeting with the public. In the experiments presented in this work, that improved fairness is achieved through medium and small costed projects. So for policymakers, this knowledge could make sure that lower costing projects are included in the election; \emph{for voters} reinforcement learning decision support has the capacity to recommend better voting strategies. Because of its optimisation it is able to identify ways to express preferences for wider social issues such that a larger share of the budget is assigned to them.

This work proposes a framework for preference formation, however there is a lot of potential to expand on this. More behaviourally accurate agent models have been proposed to improve the accuracy of agent based models \cite{Carley01121994}. It is especially important that the decision making model is the correct one for the decision making context \cite{WIJERMANS2023105850}, these can be derived using inverse reinforcement learning \cite{lee2018}, a technique by which agent models are derived from outcomes. 


\subsection*{Acknowledgements}
The authors would like to thank the reviewers of this manuscript for their invaluable feedback. This work is
funded by a UKRI Future Leaders Fellowship (MR/W009560/1): Digitally Assisted Collective Governance of
Smart City Commons–ARTIO’.
\subsection*{Data and Code Availability}
Relevant dataset and source code used for the analysis used in this paper is made available at https://github.com/TDI-Lab/FairCompromises.

\bibliographystyle{unsrt}  
\bibliography{main}  
\end{document}